\documentclass[twocolumn,preprintnumbers,amsmath,amssymb,floatfix]{revtex4}

\usepackage{graphicx}
\usepackage{dcolumn}
\usepackage{bm}

\begin{document}

\title{System size and centrality dependence of charged hadron transverse momentum spectra in Au+Au and Cu+Cu collisions at $\sqrt{s_{_{\it NN}}} =$~62.4 and 200 GeV}

\author{
%
%
B.Alver$^4$,
B.B.Back$^1$,
M.D.Baker$^2$,
M.Ballintijn$^4$,
D.S.Barton$^2$,
R.R.Betts$^6$,
R.Bindel$^7$,
W.Busza$^4$,
Z.Chai$^2$,
V.Chetluru$^6$,
E.Garc\'{\i}a$^6$,
T.Gburek$^3$,
K.Gulbrandsen$^4$,
J.Hamblen$^8$,
I.Harnarine$^6$,
C.Henderson$^4$,
D.J.Hofman$^6$,
R.S.Hollis$^6$,
R.Ho\l y\'{n}ski$^3$,
B.Holzman$^2$,
A.Iordanova$^6$,
J.L.Kane$^4$,
P.Kulinich$^4$,
C.M.Kuo$^5$,
W.Li$^4$,
W.T.Lin$^5$,
C.Loizides$^4$,
S.Manly$^8$,
A.C.Mignerey$^7$,
R.Nouicer$^2$,
A.Olszewski$^3$,
R.Pak$^2$,
C.Reed$^4$,
E.Richardson$^7$,
C.Roland$^4$,
G.Roland$^4$,
J.Sagerer$^6$,
I.Sedykh$^2$,
C.E.Smith$^6$,
M.A.Stankiewicz$^2$,
P.Steinberg$^2$,
G.S.F.Stephans$^4$,
A.Sukhanov$^2$,
A.Szostak$^2$,
M.B.Tonjes$^7$,
A.Trzupek$^3$,
G.J.van~Nieuwenhuizen$^4$,
S.S.Vaurynovich$^4$,
R.Verdier$^4$,
G.I.Veres$^4$,
P.Walters$^8$,
E.Wenger$^4$,
D.Willhelm$^7$,
F.L.H.Wolfs$^8$,
B.Wosiek$^3$,
K.Wo\'{z}niak$^3$,
S.Wyngaardt$^2$,
B.Wys\l ouch$^4$\\
\vspace{3mm}
\small
%
%
%
%
$^1$~Argonne National Laboratory, Argonne, IL 60439-4843, USA\\
$^2$~Brookhaven National Laboratory, Upton, NY 11973-5000, USA\\
$^3$~Institute of Nuclear Physics PAN, Krak\'{o}w, Poland\\
$^4$~Massachusetts Institute of Technology, Cambridge, MA 02139-4307, USA\\
$^5$~National Central University, Chung-Li, Taiwan\\
$^6$~University of Illinois at Chicago, Chicago, IL 60607-7059, USA\\
$^7$~University of Maryland, College Park, MD 20742, USA\\
$^8$~University of Rochester, Rochester, NY 14627, USA\\
}

\begin{abstract}\noindent

We present transverse momentum distributions of charged hadrons
produced in Cu+Cu collisions at $\sqrt{s_{_{\it NN}}} =$~62.4 and 200~GeV.
The spectra are measured for transverse momenta of $0.25 < p_T < 5.0$~GeV/c
at $\sqrt{s_{_{\it NN}}} =$~62.4 GeV and $0.25 < p_T < 7.0$~GeV/c at 
$\sqrt{s_{_{\it NN}}} =$~200 GeV, in a pseudo-rapidity range 
of $0.2 < \eta < 1.4$.
The nuclear modification factor $R_{\it AA}$ is calculated relative to p+p
data at both collision energies as a function of collision centrality.  
At a given collision energy and fractional cross-section, $R_{\it AA}$ is 
observed to be systematically larger in Cu+Cu collisions compared to Au+Au.  
However, for the same number of participating nucleons, $R_{AA}$ is essentially
the same in both systems over the measured range of $p_{T}$, in spite of the 
significantly different geometries of the Cu+Cu and Au+Au systems.

\vspace{3mm}
\noindent 
PACS numbers: 25.75.-q,25.75.Dw,25.75.Gz
\end{abstract}

\maketitle

The yield of charged hadrons produced at mid-rapidity in Cu+Cu collisions at 
energies of $\sqrt{s_{_{\it NN}}} =$~62.4 and 200~GeV has been measured
with the PHOBOS detector at the Relativistic Heavy Ion Collider (RHIC) at 
Brookhaven National Laboratory. The data are presented as a function of 
transverse momentum ($p_T$) and collision centrality.  The goal
of these measurements is to study the possible formation of a new form of 
matter through modification of particle production compared to 
nucleon-nucleon collisions at the same energy.

This measurement was motivated by  results from Au+Au collisions for 
$\sqrt{s_{_{\it NN}}} =$~62.4, 130 and 200~GeV.  Hadron production at these energies 
was found to be strongly suppressed relative to expectations based on 
an independent superposition of  nucleon-nucleon collisions at $p_T$ of 
2--10 GeV/c \cite{phenix_quench,phenix_highpt_npart,star_highpt_npart, phobos_highpt_npart, phobos_highpt_62}. 
The modification of high-$p_T$ hadron yields has commonly been investigated 
using the nuclear modification factor, $R_{\it AA}$, defined as 
\begin{equation}
R_{\it AA}(p_{T}) = \frac{\sigma_{pp}^{inel}}{\langle N_{\it coll} \rangle} 
                   \frac{d^2 N_{\it AA}/dp_T d\eta} {d^2 \sigma_{pp}/dp_T d\eta}.
\end{equation}
A value of $R_{\it AA} = 1$ is obtained if particle production scales
with the average number of binary nucleon-nucleon collisions, 
$\langle N_{coll} \rangle$, within a heavy-ion collision. 
Instead, for the production of charged hadrons in
central Au+Au collisions at $\sqrt{s_{_{\it NN}}} =$~200~GeV,
values of $R_{\it AA} \approx  0.2$ are observed at $p_{T}=4$~GeV/c
\cite{phenix_highpt_npart,star_highpt_npart,phobos_highpt_npart}.

Such a suppression had been predicted to occur as a consequence of the
energy loss of high-$p_T$ partons in the dense medium formed in Au+Au
collisions \cite{jet_quench_theory}.  This hypothesis is also supported by
the observed absence of this effect in deuteron-gold collisions at the
same collision energy \cite{phenix_dAu,star_dAu,brahms_dAu,phobos_dAu}.
      
\begin{table}
\caption{
 \label{table1}
 Details of the centrality classes used in this analysis.  Bins are expressed
 in terms of percentage of the total inelastic Cu+Cu cross-section.}

 \begin{ruledtabular}
 \begin{tabular}{ c c c c c }
 Centrality &
 $ \left<N_{part}^{62.4}\right>$ &  $ \left< N_{coll}^{62.4}\right>$
 & $ \left<N_{part}^{200}\right>$ &  $ \left< N_{coll}^{200}\right>$ \\
 \hline
45--50\% &  ---  & ---  &  $21  \pm 3$  & $22 \pm 4$   \\
35--45\% &  ---  & ---  &  $29  \pm 3$  & $33 \pm 5$  \\
35--40\% &  $31   \pm 3$  & $33   \pm 5$   &  ---  & ---   \\
25--35\% &  $41   \pm 3$  & $49   \pm 6$   &  $43  \pm 3$  & $56  \pm 6$  \\
15--25\% &  $59   \pm 3$  & $80   \pm 7$   &  $62  \pm 3$  & $94  \pm 8$  \\
6--15\%  &  $81   \pm 3$  & $125  \pm 9$   &  $84  \pm 3$  & $144 \pm 11$  \\
0--6\%   &  $101  \pm 3$  & $170  \pm 12$  &  $104 \pm 3$  & $197 \pm 14$ \\

\end{tabular}
\end{ruledtabular}
\end{table}

The results presented here for Cu+Cu collisions at $\sqrt{s_{_{\it NN}}}
=$~62.4 and 200~GeV bridge the gap between the Au+Au and d+Au systems, 
allowing a unique examination of the dependence of high-$p_T$ suppression 
on system size.  A careful comparison of Cu+Cu and Au+Au spectra 
to model calculations at high-$p_{T}$ can elucidate the dependence of
absorption on path length, especially as one expects a different 
distribution of path lengths for jets produced in the two systems even for the
same $N_{part}$.

\begin{figure}[ht]
\hspace{-1cm}
\centerline{
  \mbox{\includegraphics[width=7cm]{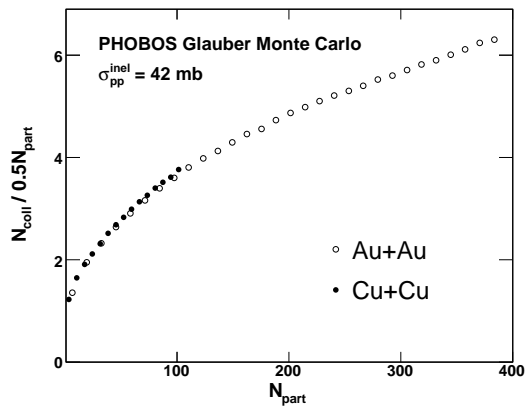}}}
\vspace{-0.3cm}
  \caption{ \label{NcollVsNpart}
  Mean number of collisions per participant pair for Au+Au (open symbols) and
  Cu+Cu (filled symbols) at $\sqrt{s_{_{\it NN}}} =$~200 GeV.}
\vspace{-0.3cm}
\end{figure}

The data were collected using the PHOBOS two-arm magnetic spectrometer. 
Details of the experimental setup can be found in \cite{phobos_nim}.
The primary event trigger used the time difference between signals in 
two sets of 10 \v Cerenkov counters, located at $4.4< | \eta | <4.9$, to select
collisions that were close to the nominal vertex position along the beam-axis.

\begin{figure}[t]
\hspace{-0.3cm}
\centerline{
  \mbox{\includegraphics[width=8.5cm]{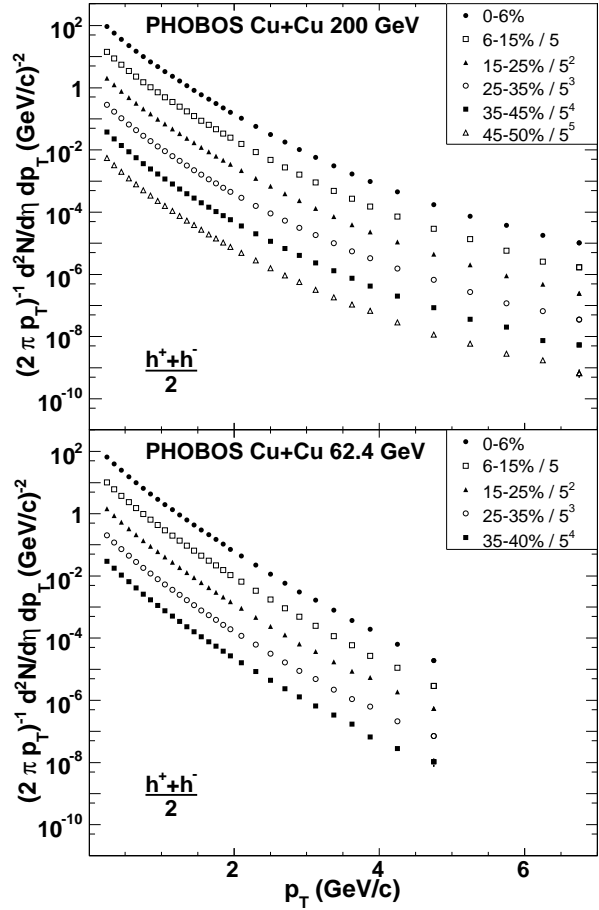}}}
\vspace{-0.3cm}
  \caption{ \label{SpectraCentBins}
  Top: Invariant yields for charged hadrons from Cu+Cu collisions at
  $\sqrt{s_{_{\it NN}}} =$ 200~GeV, in the pseudo-rapidity interval
  $0.2 < \eta < 1.4$ as a function of $p_T$ for 6 centrality bins.
  Bottom: The same for $\sqrt{s_{_{\it NN}}} =$ 62.4~GeV.  
  For clarity, consecutive bins are scaled by factors of 5.
  Statistical and systematic uncertainties are smaller than the symbol size.}
\vspace{-0.3cm}
\end{figure}

For the analysis presented here, events were divided into centrality classes
based on the total energy deposited in the octagon silicon detector, covering 
pseudo-rapidities $ |\eta |< 3.2$.  
A full detector simulation using HIJING events \cite{phobos_cent_200,hijing} 
was used to estimate $\langle N_{part} \rangle$ for each centrality class, 
and the corresponding $\langle N_{coll} \rangle$ values were obtained from 
a Monte Carlo Glauber calculation \cite{phobos_cent_200,white_paper}.
For these calculations, as well as for the determination of $R_{\it AA}$ at
62.4 and 200~GeV, we used $\sigma_{pp}^{inel} = 36 \pm 1$~and~$42 \pm 1$~mb,
respectively \cite{pdg}.  The results are listed in Table~I.  
The systematic uncertainty on these values comes primarily from
the uncertainty on our estimate of the measured fraction of the total inelastic
cross-section, which is calculated by a variety of methods in a range of 
pseudo-rapidity regions \cite{white_paper}.  The uncertainty also covers the 
range of efficiencies measured using HIJING and AMPT \cite{hijing,ampt}.  

Fig.~\ref{NcollVsNpart} demonstrates that the value of 
$\langle N_{coll} \rangle$ is essentially the same in Au+Au and Cu+Cu for the  
same value of $\langle N_{part} \rangle$.  Thus, the comparison of these two 
systems does not allow one to distinguish between scaling with participants or 
collisions.

\begin{figure*}[t]
\hspace{-1cm}
\centerline{
   \mbox{\includegraphics[width=18cm]{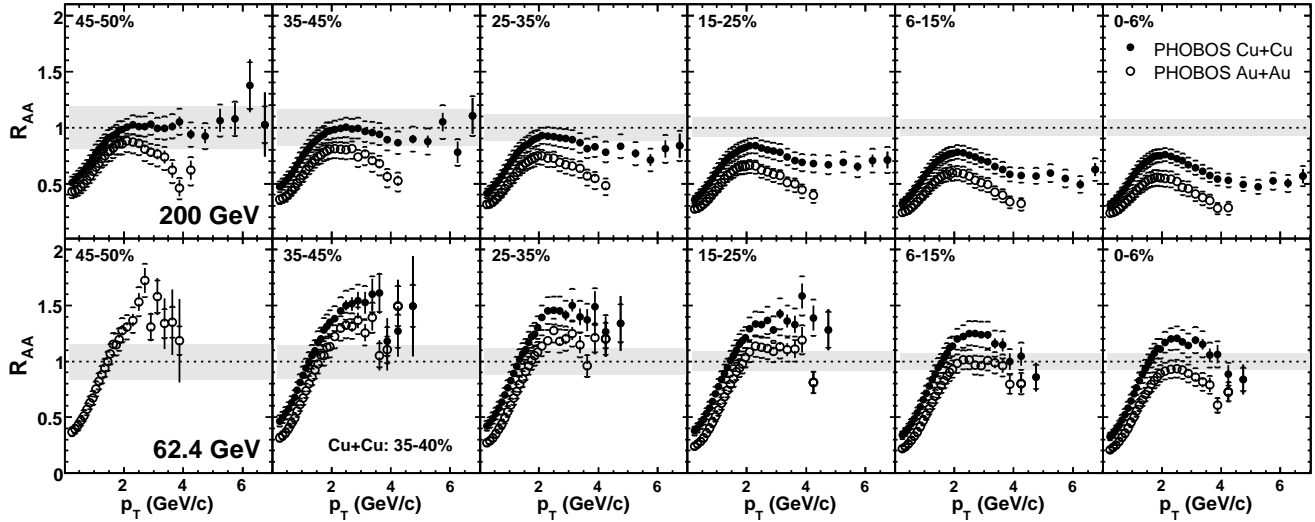}}}
\vspace{-0.3cm}
       \caption{ \label{RAAvsPt}
       Nuclear modification factor, $R_{\it AA}(p_{T})$, at mid-rapidity in bins 
       of fractional cross-section at $\sqrt{s_{_{\it NN}}} =$ 200~GeV 
       (top row) and $\sqrt{s_{_{\it NN}}} =$ 62.4~GeV (bottom row), for Cu+Cu 
       (filled symbols) and Au+Au (open symbols).  Systematic errors are shown
       with brackets (90\% C.L.).  The gray band represents the relative
       uncertainty on $\langle N_{coll} \rangle$.}
\vspace{-0.3cm}
\end{figure*}
		   
The event selection and track reconstruction procedure for this analysis 
closely followed the previously published analysis of 
Au+Au collisions at $\sqrt{s_{_{\it NN}}} = 62.4$~GeV \cite{phobos_highpt_62}. 
Events with a primary vertex position within $\pm 10$~cm of the nominal 
vertex position were selected.  Only particles traversing a full spectrometer 
arm were included in the analysis, resulting in a low transverse momentum 
cutoff at $p_T \approx 0.2$~GeV/c.

The transverse momentum distribution for each centrality bin was corrected 
separately for the geometrical acceptance of the detector, the
inefficiency of the tracking algorithm, secondary and incorrectly 
reconstructed particles,  and the distortion due to binning and momentum resolution. 
The relative importance of these corrections and their 
estimated systematic uncertainties are similar to those reported 
in the previous analysis \cite{phobos_highpt_62}. 

In Fig.~\ref{SpectraCentBins}, we present the invariant yield
of charged hadrons as a function of $p_{T}$, obtained by
averaging the yields of positive and negative hadrons. 
Data are shown for each centrality bin at both energies and are averaged over 
a pseudo-rapidity interval $0.2 < \eta < 1.4$.

The centrality evolution of $R_{\it AA}(p_{T})$ at mid-rapidity for Cu+Cu 
collisions is shown in detail in Fig.~\ref{RAAvsPt}.  
As a comparison, we also include the 
results from Au+Au collisions \cite{phobos_highpt_npart,phobos_highpt_62} 
using the same centrality binning.  As indicated, the most 
peripheral bin shown for 62.4 GeV Cu+Cu is 35-40\% central, due to the limited
efficiency of our vertex reconstruction for low multiplicity events.

At both collision energies, we notice that, in the same bin of fractional
cross-section, $R_{\it AA}$ is systematically higher in Cu+Cu compared to Au+Au.
At $\sqrt{s_{_{\it NN}}} =$ 200~GeV (top row), the Cu+Cu spectra exhibit a 
high-$p_{T}$ 
suppression, relative to binary collision scaling, ranging from approximately 
0.5 in the most central events to virtually no suppression in the most 
peripheral events.  

\begin{figure}[ht]
\hspace{-0.6cm}
\centerline{
  \mbox{\includegraphics[width=9.0cm]{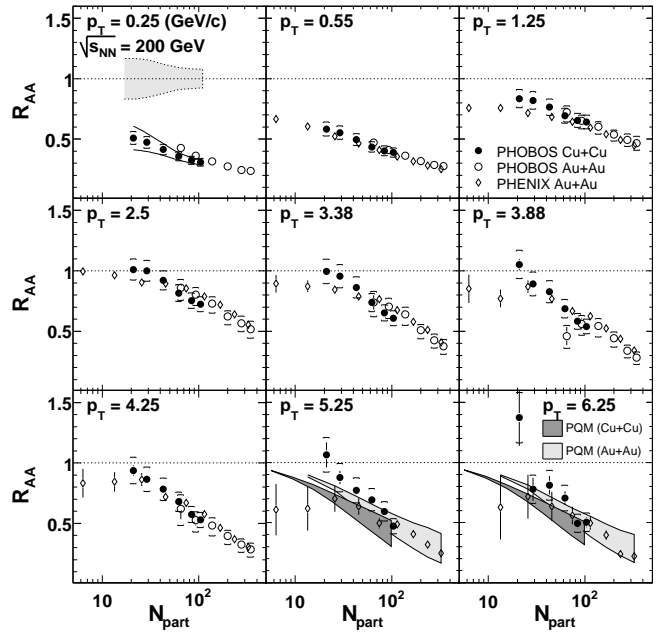}}}
\vspace{-0.3cm}
\caption{ \label{RAA200vsNpart}
  Nuclear modification factor, $R_{\it AA}$, in bins of $p_{T}$
  versus $N_{part}$ at $\sqrt{s_{_{\it NN}}} =$ 200~GeV, for
  Cu+Cu (filled symbols) and Au+Au (open symbols) 
  \cite{phobos_highpt_npart,phenix_highpt_npart}.  The gray band in the first 
  frame represents the relative uncertainty on  $\langle N_{coll} \rangle$ 
  and the solid lines show the effect of this uncertainty on the measured 
  $R_{\it AA}$.  At high $p_{T}$, bands are shown representing the predictions
  of a parton quenching model \cite{pqm}.  }
\vspace{-0.6cm}
\end{figure}

\begin{figure}[ht]
\hspace{-0.6cm}
\centerline{
  \mbox{\includegraphics[width=9cm]{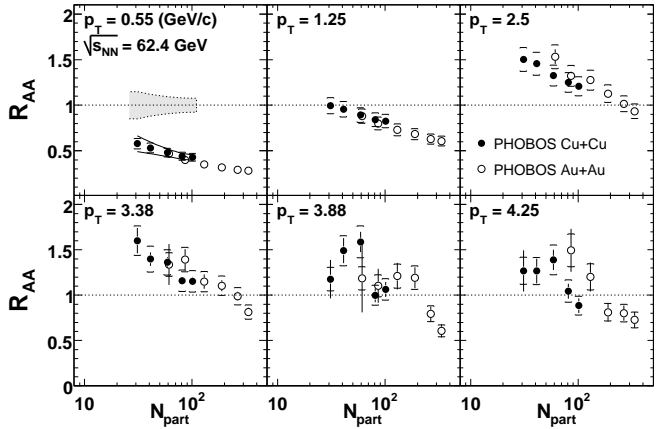}}}
\vspace{-0.3cm}
\caption{ \label{RAA62vsNpart}
  Nuclear modification factor, $R_{\it AA}$, in bins of $p_{T}$
  versus $N_{part}$ at $\sqrt{s_{_{\it NN}}} =$ 62.4~GeV, for
  Cu+Cu (filled symbols) and Au+Au (open symbols) \cite{phobos_highpt_62}.}
\vspace{-0.3cm}
\end{figure} 

For both collision energies, the spectral shape of central Cu+Cu events appears
to be very similar to the shape of peripheral Au+Au events at the same number
of participants.  This is illustrated in Fig.~\ref{RAA200vsNpart}, where
we present $R_{\it AA}$ versus the number of 
participating nucleons for both Cu+Cu and Au+Au collisions 
\cite{phobos_highpt_npart,phobos_highpt_62} in bins of $p_{T}$.  Since the 
previously published PHOBOS results did not achieve the statistics that now
allow measurements of Cu+Cu spectra out to $p_{T} =$ 7~GeV/c, we include 
results from the PHENIX collaboration \cite{phenix_highpt_npart}
for comparison.           

The result illustrated in Fig.~\ref{RAA200vsNpart} is strikingly simple.  
Over the broad range of $p_{T}$ that we measure, the 
bulk particle production seems to depend only on the size of the overlapping 
system, that is the Cu+Cu and Au+Au spectra are similar for the same number 
of participants (or binary collisions, see Fig.~\ref{NcollVsNpart}).  
Although the measured centrality ranges in Au+Au and Cu+Cu collisions have
less overlap at $\sqrt{s_{_{\it NN}}} =$ 62.4~GeV, this observation appears to 
hold at the lower energy as well (Fig.~\ref{RAA62vsNpart}).

At high-$p_{T}$, a number of predictions have been made for the Cu+Cu
system using models that successfully describe the centrality dependence
of hadron yields and back-to-back correlations in Au+Au \cite{pqm,xnwang,jet_absorption}.  One such model is the Parton Quenching Model (PQM)
\cite{pqm}, which utilizes BDMPS quenching weights \cite{bdmps} and a 
realistic collision geometry to describe partonic energy loss.  
Although the centrality evolution of the Au+Au spectra are well fit by PQM, 
our Cu+Cu results suggest that this model slightly overestimates the 
suppression in the smaller system as shown in Fig.~\ref{RAA200vsNpart},
thus this prediction does not exhibit the observed $N_{part}$ scaling.
Our conclusion is consistent with the prediction of a simple jet absorption 
model, whose only inputs are a Glauber-based collision geometry and a quadratic
dependence of absorption on path length in an expanding medium 
\cite{jet_absorption}.  
Using the absorption coefficient that describes the 200 GeV Au+Au data, this 
model gives $R_{AA}$ values which depend only on $N_{part}$, in agreement with 
our observation.   

Particle production at $p_{T} > 1$~GeV/c in heavy-ion collisions is expected
to be influenced by many effects.  These include 
$p_{T}$-broadening due to initial and final state multiple scattering (the 
`Cronin effect'), the medium-induced energy loss of fast partons, and the 
effects of collective transverse velocity fields as well as parton recombination
\cite{theory_review}.  Considering the significantly different geometries of
Au+Au and Cu+Cu collisions with the same number of participant nucleons, it 
is not obvious, {\it a priori}, that these effects should conspire to give 
similar spectra in both systems over such a large range in $p_{T}$.  
A full explanation of this observation, which appears to be a fundamental 
feature of heavy-ion collisions at these energies, may well present a challenge 
to theoretical models of heavy-ion collisions.

%
%
%
We acknowledge the generous support of the Collider-Accelerator Department.
%
This work was partially supported by U.S. DOE grants 
DE-AC02-98CH10886,
DE-FG02-93ER40802, 
DE-FC02-94ER40818,  
DE-FG02-94ER40865, 
DE-FG02-99ER41099, and
W-31-109-ENG-38, by U.S. 
NSF grants 9603486, 
0072204,            
and 0245011,        
by Polish KBN grant 1-P03B-062-27(2004-2007),
by NSC of Taiwan Contract NSC 89-2112-M-008-024, and
by Hungarian OTKA grant (F 049823).

\end{document}